# Optical skyrmions from metafibers

Tiantian He,[1,2,†] Yuan Meng,[1,†] Lele Wang,[1,2] Hongkun Zhong,[1,2] Nilo Mata-Cervera,[4] Dan Li,[1,2] Ping Yan,[1,2] Qiang Liu,[1,2] Yijie Shen[3,4,*] & Qirong Xiao[1,2,*]

1 *Department of Precision Instrument, Tsinghua University, Beijing, P.R.China, 100084*

2 *State Key Laboratory of Precision Space-time Information Sensing Technology, No.1 Qinghua Garden, Chengfu Road, Haidian District, Beijing, P.R.China, 100084*

3 *Centre for Disruptive Photonic Technologies, School of Physical and Mathematical Sciences & The Photonics Institute, Nanyang Technological University, Singapore 637371, Singapore*

4 *School of Electrical and Electronic Engineering, Nanyang Technological University, Singapore 639798, Singapore*

† Equal Contribution

* E-mail: yijie.shen@ntu.edu.sg, xiaoqirong@tsinghua.edu.cn

**Abstract:** Optical skyrmions are an emerging class of structured light with sophisticated particle-like topologies with great potential for revolutionizing modern informatics. However, the current generation of optical skyrmions involves complex or bulky systems, hindering their development of practical applications. Here, exploiting the emergent "lab-on-fiber" technology, we demonstrate the design of a metafiber-integrated photonic skyrmion generator. We not only successfully generated high-quality optical skyrmions from metafibers, but also experimentally verified their remarkable properties, such as regulability and topological stability with deep-subwavelength features beyond the diffraction limits. Our flexible and fiber-integrated optical skyrmions platform paves the avenue for future applications of topologically-enhanced remote super-resolution microscopy and super-robust information transfer.

**Introduction**

Skyrmions, as a kind of topological quasiparticles with sophisticated spin textures, originate from particle and solid-state physics as novel information carriers in local data storage[1-4]. Following their magnetic counterparts, skyrmions have recently been constructed in photonics realm as structured electromagnetic fields, namely optical skyrmions[5]. They have emerged as new candidates to revolutionize long-range topological information transfers and light-matter interactions by their abundant intriguing physical attributes[5-7]. However, the current optical skyrmions generators necessitate complex and bulk systems hindering further applications, given the challenges to generate intricate structured light fields with sophisticated polarization distributions via conventional methods. The initial formation of optical skyrmions was constructed by harnessing evanescent electromagnetic fields via surface plasma[8-10]. Soon after, optical skyrmions with tailored topological properties have also been generated using free-space optical vector fields in complex setups[11-17]. Moreover, the free-space skyrmionic beams with spin-orbit coupling and tunable topological textures[18-20] showcase unique features such as topological stability or resilience[21-25], exhibiting great potential as robust information carriers for new generation optical information networks. However, current experimental demonstration of such optical skyrmionic beams requires the use of bulky spatial light modulators[12,16,17], or cascaded integral optical elements[26]. A compact, flexible and integrated micro-generator of optical skyrmions, which is thus highly desired to unlock practical applications, is still elusive to the best of our knowledge.

As a cornerstone underpinnding modern communication systems, optical fibers[27] have established themselves as the prime choice for long-distance and high-capacity data transmission[28] thanks to their distinctive high aspect ratio, great flexibility, broad bandwidth, and immunity to electromagnetic interference. Besides, optical fibers also exhibit tremendous potential to realize optical device integration with decent miniaturization, flexibility, and lightweight. Driven by the maturation in "lab-on-fiber" technology[29], the integration of multifunctional nanostructures to fiber facets has become a hotspot thanks to its high degrees of design freedom, integration handiness, and versatile functionalities. Current approaches typically leverage 3D printing[30-32] to directly fabricate the same-scaled-down structured devices on the fiber tip. However, this strategy suffers the shortcomings of limited processing precision, restricted numerical aperture, and typically single device functionality, rendering them incapable of realizing the desired skyrmion excitation that demands fine subwavelength structure resolution to enable intrincated light field modulations. In contrast, metasurfaces[33-35] consisting of subwavelength nanoscatterers have proven paradigms to offer precise, powerful, and multi-dimensional control over the amplitude, phase, and polarization of light[36], outstanding itself as a remarkable choice to facilitate the miniaturization and creation of high-dimensional structured light. So far, metadevices have shown generators of conventional structured light[37-39], but has not yet been applied to topological optical skyrmions.

Here, we demonstrate the first on-demand generation of optical skyrmions in fibers-based metadevice, to the best of our knowledge. By judiciously designing the polarization-dependent metasurface resting on a fiber-tip, a zero-order Bessel beam and a first-order Bessel beam carrying orbital angular momentum (OAM) with high NA up to 0.8 are excited in orthogonally polarized states, respectively. The spatial entanglement of the two beams occurs in the near-field of fiber facet, inventively exciting the skyrmion. It is worth mentioning that our proposed device can be dynamically switched between "on" and "off", i.e. it can be tuned for non-skyrmion and skyrmion states generation by adjusting the polarization angle of input/output light with high flexibility. We also applied a quarter-wave plate (QWP) to transform the textures between bimeron and skyrmion, proving a good tolopogical propoties of the excited skyrmions. In addition, we experimentally confirm the effective high-quality excitation and modulation of bimeron and skyrmion, achieving skyrmion number up to 0.97, covering a nearly-full map on Poincaré Sphere. Moreover, subwavelength-scale variation of the longitudinal component of the polarized Stokes parameters of the skyrmion has been experimentally verified to be down to $\sim\lambda/5$ firstly, paving the way for future ultra-high-density and ultra-stable information storage. In conclusion, our metafiber skyrmion scheme significantly increase flexibility and integration, holding great promise for exploring topological photonic physics, optical communications, and super-resolution microscopy.

**Results**
**Concept.** Optical skyrmion is a quasiparticle state of light with distinctive polarization and topological protections[5], and it could be constructed by a group of vector beams. The construction and propoties of optical skyrmions based on the Bessel beams is schematically shown in the insert box of Fig. 1. The expansion basis can be either LG modes or Bessel modes, being Bessel modes the ones chosen in this paper. By superimposing two modulated orthogonally polarized Bessel beams (BB) : zero-order BB and first-order BB carrying orbital angular momentum (OAM)[7], in right-handed and left-handed circular polarizations (RCP and LCP) respectively, as shown in Fig. 1(a), a kind of skyrmion can be constructed. The resultant spatially-varying vector pattern will fulfill a skyrmion mapping, see Fig. 1(b), i.e. the mapping of all states from a parametric sphere to a localized plane[18]. To visualize this mapping, we utilize hue-lightness (HL) colors to label states of polarization. Each polarization state corresponds to a point on the Poincaré sphere and is represented by a unit normal vector ($s_x$, $s_y$, $s_z$) at the surface of the sphere, called Stokes vector. We use the lightness color (from black to white) to represent the longitudinal component $s_z$ (from -1 to 1 or down to up) and use hue color to visualize the azimuth of transverse ($s_x$, $s_y$) components. Thus, the skyrmions can be identified by the HL-colored map of corresponding vector fields. For instance, in the HL map of the Bessel-based skyrmionic vector beam (Fig. 1(c)), there is a full HL color area at the center, which shows a stereographic mapping to the Poincaré sphere (Fig. 1(d)).

The schematic diagram of the metafiber-based skyrmion generators is illustrated in Fig. 1, consisting of a polarization-maintaining single-mode fiber (PSF), an expansion fiber, and a single layer metasurface on the fiber-tip. A detailed description on the expanding fiber and the light beam diffusion are elaborated in Supplementary Note 2. We designed the metasurface to independently modulate the two orthogonal polarized light when passing through the meta-tips. The expansion basis can be either LG modes or Bessel modes, being Bessel modes the ones chosen in this paper. Here, Hence, the beams are tightly focused within a $15\lambda$ nondiffraction distance off the fiber facet in free space and entangled organically to generate skyrmion in subwavelength scale. The intensity and Stokes vector distribution of the whole light profile are shown in the right-bottom of Fig. 1, and a corresponding hedgehog-like Stokes vector configuration is illustrated enlarged. As we can see, the vector texture points up and down in the center and the edge of the confinement region of the localized skyrmion.

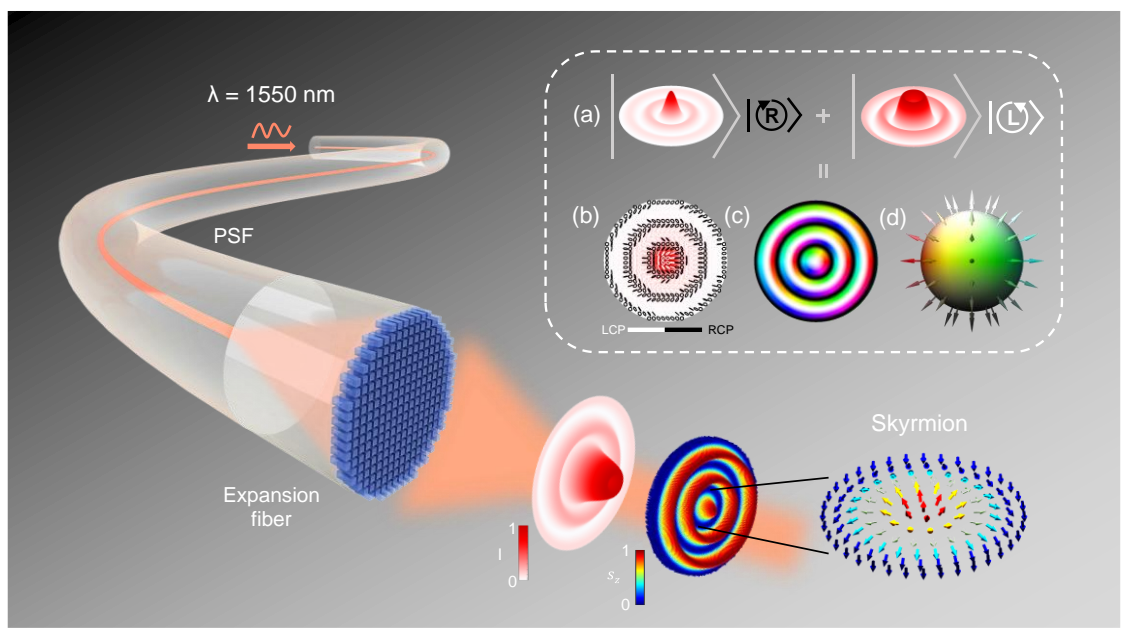

Fig. 1 The schematic diagram of the skyrmions' excitation from metafiber. The intensity and stokes vector distribution of the excitation light are shown (The vector of one whole skyrmion is detailed). Insert box: (a) The construction principle of skyrmion, composed by zero-order BB and first-order BB under orthogonal circular polarizations. The properties of the excited skyrmion including (b) polarization distribution and (c) spin distribution with (d) a HL-colored Poincaré sphere.

**Simulation and manipulation of skyrmions.**

To construct skyrmions, two different beams should be judiciously modulated on an orthogonally polarized basis, one of which carries OAM. The polarization basis can be chosen as, for instance the $x$ and $y$ linear polarizations, as well as the LCP and RCP basis, both of which can be interchanged by applying a quarter wave plate (QWP). In this study, two orthogonal linearly polarized states are chosen as basis. We demonstrated to modulate the zero-order BB in $x$-polarization and the first-order Bessel light in $y$-polarization respectively to excite skyrmion. The metasurface is made of silicon nanoantennas on top of quartz substrate, with a lattice period of 700 nm for meta-units. The size of the Si antennas ranges from 250 nm to 550 nm to ensure a full $2\pi$ phase coverage (Supplementary Note 1). The 50 μm-diametered metasurface contains more than 4,000 nanoantennas in total. The propagation phase concept is used for the design of dual-Bessel beams with orthogonal polarizations. The $J_0$ beam can be generated by printing an axicon phase profile on a plane wave, and introducing a vortex phase into $J_0$ yields the vortex BB $J_1$. Indeed, as the phase of the output light field of fiber is curved rather than uniform, it is necessary to superimpose a correction phase onto the original phase profile to properly compensate the divergint wavefront of the expansion fiber facet. By specific design of metasurface (details in the Supplementary Note 1), dual-order BBs under different polarizations can be achieved.

Theoretically, the transverse intensity profiles of the BBs remain constant as the propagation distance $z$ increases showing diffractionless feature. However, these ideal BBs require an infinite amount of aperture and energy as well, making it not practically possible to excite ideal BBs using infinite sized optical devices. Real devices with finite aperture limit the non-diffracting behaviour of quasi-BBs to maintain within a certain finite distance as proven in previous works[36]. For the topological quasi-particles, smaller skyrmion size will instead lead to higher information storage density. Therefore, Bessel beams with NA up to 0.8 are chosen here to exploit the potential of sub-wavelength skyrmion generation in metafibers. A longitudinal slice of the intensity of zero-order BB $J_0$ and first-order BB $J_1$ are illustrated in the first row of Fig. 2(a). The intensity and phase profiles at the transverse plane highlighted with the dashed white lines are shown in the bottom two rows of Fig. 2(a). The left two columns correspond to the ideal profiles and the third and fourth to the simulated ones, showing a excellent consistency. Herein, the FWHM of the $J_0$ is defined as the distance between two points at half of the maxima intensity of the center bright spot. Similarly, the FWHM of $J_1$ is defined as twice the distance from the dark spot center to the point at its closest ring with the half-maximal intensity. The ideal FWHM can be derived as[36]: $\text{FWHM}_{J_0} = 2.25/k_r = 0.358\lambda/\text{NA}_1$, $\text{FWHM}_{J_1} = 1.832/k_r = 0.292\lambda/\text{NA}_2$, where $k_r = \text{NA} \cdot 2\pi/\lambda$. By choosing the $\text{NA} = \text{NA}_1 = \text{NA}_2 = 0.8$, the simulated FWHM values of $J_0$ and $J_1$ are 0.64 μm and 0.57 μm respectively, showing a great consistence with the ideal FWHMs of 0.69 μm and 0.57 μm. When the light beam $J_0$ is superposed with $J_1$ in orthogonal polarizations, the skyrmionic topological states are excited.

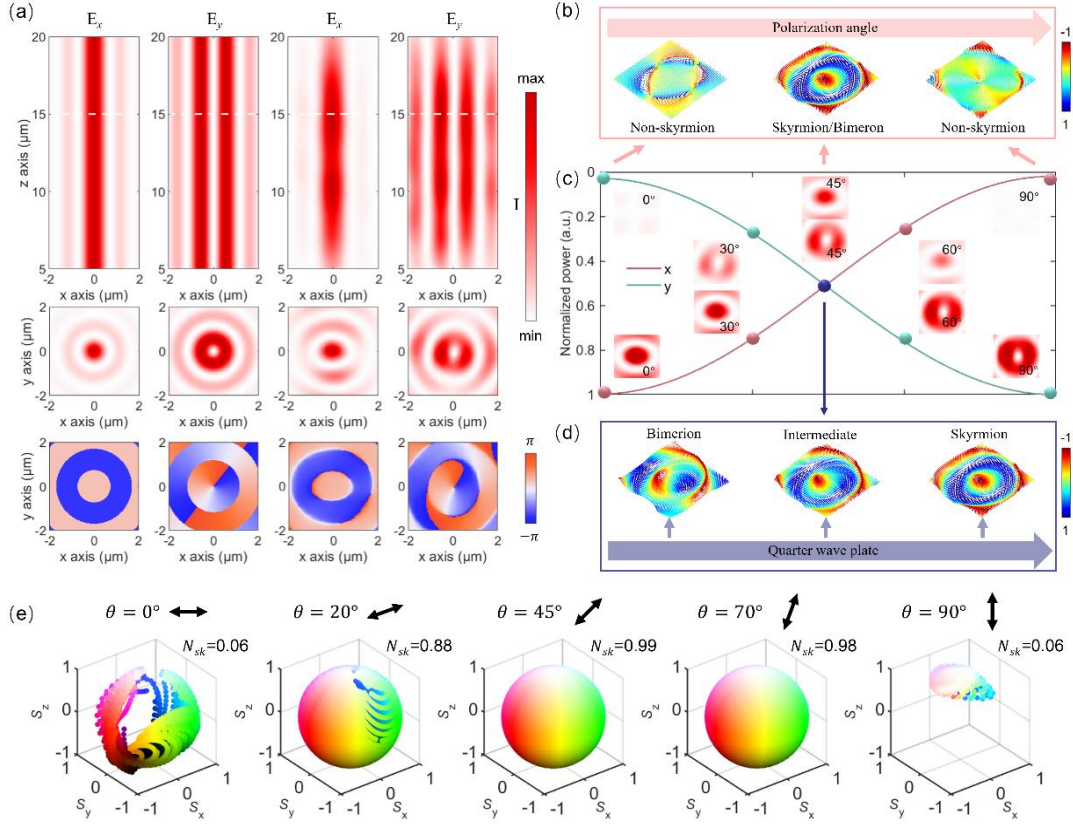

Fig. 2 (a) Ideal and simulated light field distributions of skyrmion components of zeroth-order BBs $J_0$ and first-order BBs $J_1$. The first and second columns are ideal profiles and the third and fourth are simulated profiles. The intensity profiles at the longitudinal plane $xz$ ($y = 0$), and the intensity and phase at the transverse $xy$ plane ($z$ is marked with white lines in the first row) are illustrated from top to bottom. (b) The manipulation of skyrmion/non-skyrmion states via polarization angles $\theta$ with the vector distribution of major states. (c) Normalized power curves with respect to the polarization angle. The intensity profiles of $J_0$ and $J_1$ of different polarization angles are inserted and marked in lines with dots. (d) The manipulation of topological transformation between different quasiparticles via quarter wave plate. (e) The map on the Poincaré Sphere in the transformation between skyrmion and non-skyrmion in different polarization angle $\theta$ in simulation.

The continuous modulation is a crucial property in the device, which not only facilitates a deeper comprehension of topological transformations, but also substantially elevates the degree of integration in specific applications like multilevel information encoding. The modulation of our proposed metafiber is shown in Figs. 2(b)–2(d). On the one hand, by tuning the polarization angle of source, a smooth evolution from non-skyrmion and skyrmion can be achieved, equivalent to the evolution from non-OAM to OAM states. To clearly demonstrate this continuous transformation, a quantitative curve is presented in Fig. 2 (c). The absolute value of the angle between source polarization direction and $x$-axis is defined as $\theta$ in degrees. The power is calculated integrating the intensity over the entire transverse cross-section. The total power of the source $P_t$ remains unchanged in this transformation. The pink and green lines represent the power in $x$ and $y$ polarization respectively, with the intensity profiles embedded in Fig. 2(c) for different input polarization angles 0°, 30°, 45°, 60°, and 90°. It is evident that the power of the $J_0$ progressively diminishes as the polarization angle increases, whereas $J_1$ steadily increases, due to the near-unity transmission of the metasurface for any polarization input. The trend can be quantitatively fitted as $P_{x,input} = P_t \times \cos^2(\theta)$ and $P_{y,input} = P_t \times \sin^2(\theta)$. Stokes vector distributions of typical output states following a QWP, which transitions from linear to circular polarization, are shown in Fig. 2(b). When $\theta = 45°$, the vector shows a hedgehog skyrmion distribution with a vortex around the center. On the other hand, the skyrmion vector textures can be modified through topologically preserving smooth transformations. In order to verify this property, we introduce a QWP to change to different quasiparticles' textures. Typical vector distributions are shown in Fig. 2(d). Specifically, a bimeron is achieved by line polarization modulation, and skyrmionic states are obtained when placing a QWP with the fast axis at 45° with respect to the *x*-axis, and remaining bimeron states when using a waveplate with a fast axis parallel to the *x*-axis. When the fast axis of the QWP aligns with the *x*-axis, the vector distribution of the Bimeron state exhibits a texture composed of two half-merons with opposite polarities. By rotating the QWP from 0° to 45°, the intermediate states appear.

The Poincaré Sphere provides a graphical visualization of the Stokes, allowing a clearer representation of the modulated vector transformations. The coordinate axes of the Poincaré sphere are normalized $s_x$, $s_y$, $s_z$. For the polarization angle regulation, the skyrmion maps on the Poincaré sphere are plotted in Fig. 2(e), with the polarization angle $\theta$ and the source electric field oscillation directions are also labeled. When the polarization angle is 0°, the points on the Poincaré sphere are very sparse, and the skyrmion number is 0.06, showing a non-skyrmion state. As the polarization angles increases, the vector distributions gradually spread until they fully cover the Poincaré sphere. When the polarization angle is equal to 45°, the Poincaré sphere has dots distributed all over the sphere and contains a full HL color, demonstrating a perfect skyrmion with a skyrmion number of 0.99. In addition, our proposed device can modulate a high quality skyrmion at angles between 20-80 degrees, which means that the topological fields can be generated without the need of very precise adjustments, making it very valuable for practical applications. Meanwhile, the texture transformations can also be characterized by means of the Poincaré sphere. The interconversion of skyrmion and bimeron can be realized by QWP, whose effect is equivalent to the transformation of the orthogonal base of polarization, which also corresponds to a coordinates transformation on the Poincaré sphere. To take a special case, when the angle between the fast-axis of a QWP and the *x*-axis is 45°, the coordinate system of the Poincaré sphere ($s_x,s_y,s_z$) changes to ($s_z,s_x,s_y$). This coordinate transformation does not affect the topology as well as the skyrmion number, but causes a distribution transformation of the Stokes parameters, as specified in Methods. The regulation of the skyrmion state and the bimeron state with detailed $\theta$ are in Supplementary Note 3, and the skyrmion number along with $\theta$ is also plotted in supplementary Note 4. Besides, a favorable broadband characteristic beyond 200 nm are also realized in our proposed design, as detailed in Supplementary Note 5.

**Experimental results.**
The direct attaching method is used to produce the proposed devices. Firstly, the metasurface is manufactured by standard electron beam lithography (EBL) and lift-off process (details are in Methods). Meanwhile, the PSF (PM1550-HP) is attached to a fiber holder for stability, with the cross sections polished flat together. Then the glue is dripped between the fiber and the metasurface. For precisely center them, a homemade amplified light system is constructed to achieve a magnification of 100× (Detailed in Methods). The thickness of the glue is precisely controlled by the micromotorized displacement stage, which first focuses the focal plane on the fiber facet, then moves the micromotorized displacement stage in aim distance, and finally moves the metasurface to make a clear imaging. Finally, the proposed device can be obtained by curing the adhesive with a UV lamp. A physical picture of the integrated device and a oblique scanning electron microscope (SEM) of the local metasurface are depicted in Fig. 4(a), with a detailed chart inserted.

To validate the properties of the excited skyrmion, the complex amplitudes of the light field in $x$ and $y$ polarization are required, which can be reconstructed by the phase recovery method (details are in the supplementary materials). Therefore, the experimental setup is built (specified in the Methods as well as illustrated in supplementary materials), including an amplified path and an interference parallel light path. Then, the intensities of the interfered light fields and the reference plane fields are recorded. Consequently, through the phase recovery method, the complex amplitude of the object field under two orthogonal polarizations are obtained, as shown in the bottom of Fig. 4(b), while the upper row shows simulation results. The experimental $x$-polarized intensity distribution shows a bright central spot surrounded by a dark ring corresponding to the first zero, and the phase is nearly uniform in the beam's central region. The y-polarized intensity pattern has a typical doughnut shape with a dark dot in the center surrounded by a bright ring, while the resulting helical phasefront presents a point singularity in the origin, showing a good agreement between the experimental results and the simulations.

By mapping $J_0$ in the polarization state |H⟩ and $J_1$ in |V⟩, the bimeron state is realized. Afterwards, using the half wave plate to convert the polarization of $J_0$ as |R⟩ and $J_1$ as |L⟩, the skyrmion states can be realized as well. The spin distributions of the skyrmion and bimeron state are shown on the left and right sides of Fig. 4(c), respectively. From the lower left corner, skyrmion spin distribution shows a white central, indicating a left-handed circular polarization, surrounded by a black circle representing a right-handed circular polarization. A colored ring appearing between the black and white regions reveals the coverage of all the intermediate polarization states in the Poincaré sphere, and the colors changing counterclockwise according to the HSV. On the other hand, the spin distribution of the bimeron state in the right corner exhibits black and white regions on the left and right sides, and varying colors in the middle, showing two merons with opposite polarity. Additionally, the density and polarization distributions of the realized skyrmion are shown in Fig. 4(d). After the calculation of the topological properties as detailed in Methods, we find that the skyrmion number in the experiment is 0.97, which is absolutely close to the value in the simulations (0.99), proving the generation of a high quality skyrmion from the metafiber.

Furthermore, sub-wavelength characteristics of the polarization Stokes parameters are also realized in our proposed meta-skyrmion emitter. The stokes vector distributions of the skyrmion and bimeron states of experiments are illustrated in Fig. 4(e), and the corresponding $S_z$ curves along the red line are also inserted (The calculated Stokes parameters are detailed in Supplementary Note 10). The simulation and experimental results are orange and ink-green, with the original data presented with increased transparency and its absolute value plotted without transparency. The polarization inversion width is defined through the point at which the Stokes parameter $S_z$ changes its sign and reaches half the maximum value of $S_z$. It can be seen that both skyrmion and bimeron reverse the polarization state on within a distance of λ/5.1 and λ/4.6 respectively, which is much smaller than the diffraction limit λ/2 and the FWHM of the beams. Therefore, in agreement with our theoretical proposal, we experimentally verify for the first time the sub-wavelength features of the generated Stokes topological fields, which can be of great importance for applications where miniaturization and integration is required.

We can also see reasonable discrepancies in the resulting skyrmion textures between the experiments and the simulations, mainly due to errors in the measurement of the electric field caused by several aspects. Firstly, the size of the generated beams is extremely small with respect to the pixel size of the camera (5.2 μm), leading to a sparse sampling although the captured beams are magnified by 100×. The insufficient resolution may result in the loss of points, causing a slight difference in the intensity profiles. Secondly, the process of complex amplitude recovery is sensitive to misalignements between the two arms of the interferometers, which include the propagation directions as well as polarizers' directions in the two optical paths cannot be perfectly parallel, yielding small errors in the reconstructed complex field. Thirdly, the accuracy of the processing methods could cause discrepancies with respect to the ideal case. After computationally extensive simulations and experiments, taking into account the constrained processing conditions, the phase shift of $2\pi$, and the optimal illumination scheme, an antenna period of 700 nm with dimensions ranging from 250 to 550 nm was determined. However, deviations are inevitable during metasurface fabrication and experiments. Still, the experimental results shows the great robustness of the device under the different kinds of imperfections. A detailed error analysis the metasurface is presented in the Supplementary Note 6, demonstrating an excellent robustness of our proposed device.

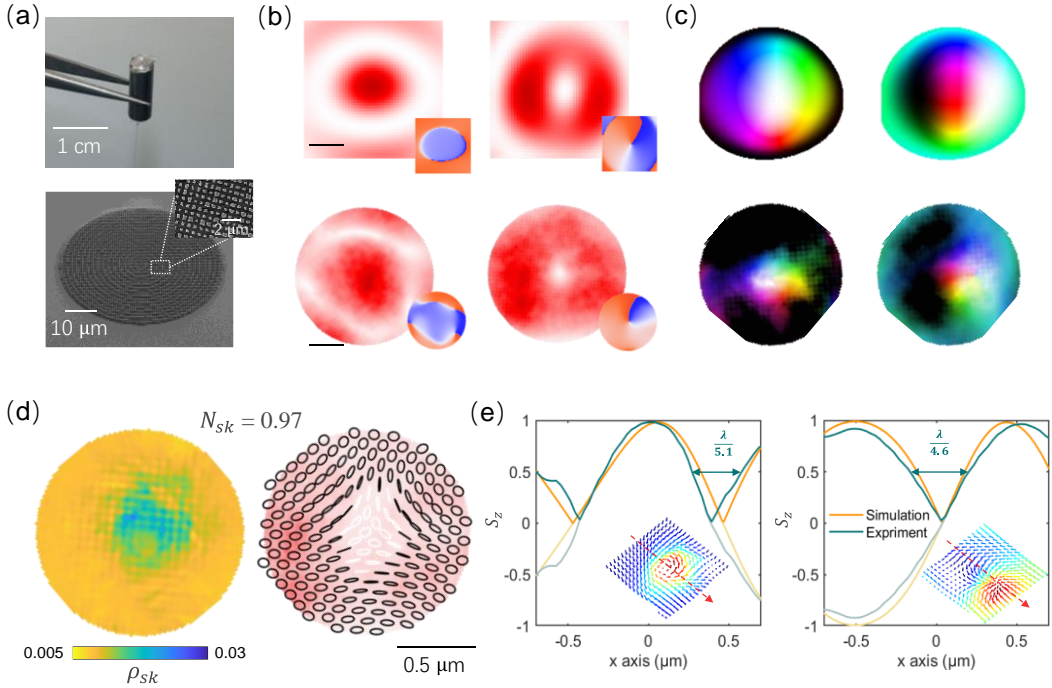

Fig. 3 (a) The physical picture of the whole device (upper). The SEM of the whole metasurface (lower) and the top view of local metasurface (Inserted box). (b) The complex amplitude profiles of the simulation (upper) and experimental (lower), respectively. (c) The spin distributions of skyrmion (left) and bimeron (right) of the simulation (upper) and experiment (lower), respectively. (d) The skyrmion density and polarization distribution of the skyrmion state in experiments. (e) The normalized Stokes parameter and its absolute value versus $x$ of skyrmion (left) and bimeron (right) states, with the vector distributions of experiment results inserted. Scale bar: 500nm.

**Discussion**

In conclusion, we have demonstrated the metafibers as very compact and flexible generators of optical skyrmion. An efficient and simple scheme is proposed to excite skyrmion in free space by carefully designing the metafiber. The metasurface is endowed by two beams in orthogonally polarized states independently, one of which carries OAM. Either Bessel mode or LG mode can be chosen as the basis. Bessel beams, specifically the zero-order and the first-order Bessel beams, are chosen for demonstration. By tuning the polarization angle of source, a flexible continuum modulation between the skyrmion and non-skyrmion states are realized. The transformation of textures between the skyrmion and bimeron quasiparticle states is also verificated by a QWP placed after the metafiber. We also experimentally demonstrate the features of the metafiber and validate the qualy of the excited skyrmions, exhibiting a high skyrmion number up to 0.97. By using a polarizer and a QWP in free space, the non-skyrmion, skyrmion and bimeron states are all realized. It is worth mentioning that the sub-wavelength fashion of polarized Stokes parameters is proposed and experimentally verified for the first time. Compared to previous methods of spins[9] and magnetic fields[14], our proposed scheme can be detected easier, e.g., by laminar measurements, and can be extended more broadly to kinds of applications, e.g., optical tweezers, imaging, etc.. In addition, our device demonstrates remarkable stability to generate high quality skyrmions in a broad range of input polarization angles. In general OAM devices, the light field distribution becomes elliptical[36] instead of circular when the NA becomes high, disturbing the quality of the generated OAM modes, but in our case, the stokes textures generated by our proposed device carry a NA-invariant skyrmion number. In addition, the metafiber could also be designed using other methods such as geometric phase to directly modulate the LCP and RCP polarizations.

In brief, it is a very promising direction to control more topological states in subwavelength meta-devices with ultrasmall volume for achieving ultra-high density information encoding. The metafiber design could also be extended for generating more diversified and topological quasiparticle states by superimposing sophisticated beams, for instance the higher-order skyrmions, multiskyrmions, and 3D hopfions. Besides, the device size and processing time of metafiber can be further reduced through direct processing in the fiber facet[37,40], which is the trend in future. To enhance the compactness of the device , the polarizer can be replaced by performing polarization control in the all-fiber source and the QWP can be optimized by adding a rotating metasurface[41] or directly using a two-layer metasurface[42,43], which is the next step of our ongoing work. Moreover, by combining phase change materials[44-46], two-dimensional materials[47], or liquid crystals[48,49] with metasurfaces, the tunability between different quasiparticle states can be further enhanced. Therefore, the proposed of the metafiber skyrmion generator could strongly impact the development of new-generation information technologies. Given the versatility of meta-structures, the proposed design can be further extended to generate various optical skyrmions, also synergizing the unique advantages of optical fibers. The current fiber communication networks can be upgraded into metafiber networks where skyrmions are the novel information carriers for larger-capacity information transfers. In addition, in contrast to conventional light waves, optical skyrmions can provide unconditional stability and resilience protected by their topologies against perturbations of disorders or turbulence in information channels, promising their usage of upgrading ultra-robust communication networks.

## Methods

**Metasurface fabrication.** After the meticulous design of the metasurface, it was manufactured using the following processing procedure (Supplementary Note 7), with flow chart shown as supplementary Fig. S4. First, a 900nm thick amorphous silicon is grown on a quartz substrate using

Electron Beam Physical Vapor Deposition (EB-PVD). Then, the amorphous silicon layer is precisely etched using Electron Beam Lithography (EBL) to create the required structure, which includes a lattice period of 700 nm and antenna dimensions ranging from 250 nm to 550 nm. After etching, a chromium (Cr) layer is deposited on the amorphous silicon surface using Physical Vapor Deposition (PVD) technology, followed by a carefully designed lift-off process to remove excess chromium. Finally, the target metasurface is achieved through Inductively Coupled Plasma Etching (ICP) and the removal of the Cr layer. The fabrication procedure's chart is illustrated in Fig. S3.

**Experiment Setup.** A homemade optical measurement system is constructed to observe the performance of the metafiber skyrmion's generator, including magnification and coherent optical path. The diagram of the experimental setup is shown in Figure S6. A homemade fiber laser operates at a wavelength of 1550 nm and outputs a power of 0-300 mW. Through a polarization beam splitter, light with two polarizations is directed into the PSF of the light field amplification path and the coherent light path, respectively. The polarization controller attached to the laser output controls the intensity of light from the laser output to the two pathways. In the amplification path, one end output of PBS couples with metafiber to excite the skyrmions. As the lateral output spot of the skyrmions is on the sub-micron scale, and the size of a single pixel of the CCD (CinCam CMOS-1201) is 5.2 μm × 5.2 μm, a 100× near-infrared objective lens is required for magnification. The objective lens is fixed on a motorized micrometer translation stage to move accurately. The polarizer is used to separate the light of different polarizations. In the coherent light path, the fiber is attached to a collimating mirror for light expansion, and light turns through a reflector. The size of light after collimation is close to the size of the metasurface. Then, the light from two light paths is merged through the beam splitter and interferes. The interfered light fields in x and y polarizations could be obtained by setting the directions of the two polarizers the same for x and y, respectively.

**Characterizing topology of skyrmions.**
Topological properties of a skyrmionic configuration can be characterized by the skyrmion number[14]:

$$N_{sk} = \frac{1}{4\pi} \iint_\sigma \rho_{sk} \, dx dy \qquad (1)$$

where the skyrmion density is given by:

$$\rho_{sk} = \mathbf{s} \cdot \left( \frac{\partial \mathbf{s}}{\partial x} \times \frac{\partial \mathbf{s}}{\partial y} \right) \qquad (2)$$

where $\mathbf{s}(x,y) = [s_x(x,y), s_y(x,y), s_z(x,y)]$ ($s_x^2 + s_y^2 + s_z^2 = 1$) represents the Stokes vector field to construct a skyrmion and is the region considered to confine the skyrmion. The skyrmion number is an integer counting how many times the vector $\mathbf{s}(x,y) = \mathbf{s}(r\cos\theta, r\sin\theta)$ wraps around the parametric unit sphere, i.e. the Poincaré sphere. For mapping to the unit sphere, the vector can be given by $\mathbf{s} = (\cos\alpha(\theta)\sin\beta(r), \sin\alpha(\theta)\sin\beta(r), \cos\beta(r))$. The skyrmion number can be separated into two integers: the polarity, $p = \frac{1}{2}[\cos\beta(r)]_{r=0}^{r=r_\sigma}$, means that the vector direction is down (up) at center $r = 0$ and up (down) at the boundary $r \to r_\sigma$ for $p = 1$ ($p = -1$), and the vorticity, $m = \frac{1}{2\pi}[\alpha(\theta)]_{\theta=0}^{\theta=2\pi}$, controls distribution of the transverse field components. In the case of a helical distribution, an initial phase $\gamma$ should be added, $\alpha(\theta) = m\theta + \gamma$. For continuous regulation, the Jones matrix expression for a quarter-wave film rotated at $\theta$ is given by $\begin{bmatrix} 1 - i\cos 2\theta & -i\sin 2\theta \\ -i\sin 2\theta & 1 + i\cos 2\theta \end{bmatrix}$. It is a unitary matrix, meaning that the QWP can be regarded as a rotation of the coordinate. Mathematically, the skyrmion number has been proven to be independent of the orthogonal coordinate system used[50]. From a topological point of view, this transformation is a smooth deformation that does not change the topology and therefore does not affect the skyrmion number.